\documentclass[referee]{raa}            

\usepackage{graphicx,times}             
\usepackage{epstopdf}
\usepackage[free-standing-units]{siunitx}

\begin{document}

   \title{Investigating the Efficiency of the Beijing Faint Object Spectrograph and Camera (BFOSC) of the Xinglong 2.16-m Reflector
}

   \volnopage{Vol.0 (20xx) No.0, 000--000}      
   \setcounter{page}{1}          

   \author{Yong Zhao
      \inst{1,2}
   \and Zhou Fan
      \inst{1}
   \and Juanjuan Ren
      \inst{1}
   \and Liang Ge
      \inst{1}
   \and Xiaoming Zhang
      \inst{1}
   \and Hongbin Li
      \inst{1}   
   \and Huijuan Wang
      \inst{1}
    \and Jianfeng Wang
      \inst{1}  
    \and Peng Qiu 
      \inst{1,2}
    \and Xiaojun Jiang
      \inst{1,3}
   }

   \institute{Key Laboratory of Optical Astronomy, National Astronomical Observatories, Chinese 
 Academy of Sciences, Beijing 100012, China; {\it zhaoyong@nao.cas.cn}\\
        \and
            University of Chinese Academy of Sciences, Beijing 100049, China\\
            \and School of Astronomy and Space Science, University of Chinese Academy of Sciences, Beijing 100049, China\\
   }

   \date{Received~~2009 month day; accepted~~2009~~month day}

\abstract{
The Beijing Faint Object Spectrograph and Camera (BFOSC) is one of the most important instruments of the 2.16-m telescope of the Xinglong Observatory. Every year there are $\sim20$ SCI-papers published based on the observational data of this telescope. In this work, we have systemically measured the total efficiency of the BFOSC of the 2.16-m reflector, based on the observations of two ESO flux standard stars. We have obtained the total efficiencies of the BFOSC instrument of different grisms with various slit widths in almost all ranges, and analysed the factors which effect the efficiency of telescope and spectrograph. For the astronomical observers, the result will be useful for them to select a suitable slit width, depending on their scientific goals and weather conditions during the observation; For the technicians, the result will help them systemically find out the real efficiency of telescope and spectrograph, and further to improve the total efficiency and observing capacity of the telescope technically.
}

   \authorrunning{Y. Zhao et al.}            
   \titlerunning{Investigating the Efficiency of the BFOSC }  

   \maketitle

%
%
\section{Introduction}           
\label{sect:intro}

The Xinglong 2.16-m reflector is an English equatorial mount telescope at Xinglong Observatory, with the effective aperture of 2.16 meters, and the focal ratio of f/9 at Cassegrain focus (Su et al.~\cite{{1989SIC..11...1187}}). It is the first 2-meter class astronomical telescope in China. The 2.16-m reflector is equipped with three main instruments currently: (1) the BFOSC (Beijing Faint Object Spectrograph and Camera), which is used for imediate and low-resolution (R $\sim$ 500 -- 2000) spectroscopy; (2) the OMR (the spectrograph made by Optomechanics Research Inc, in Tucson, Arizona), which is used for low resolution spectroscopy, with the similar spectral resolutions to those of the BFOSC; (3) the HRS (High Resolution Spectrograph), which is used for fiber-fed high-resolution (R $\sim$ 30000 -- 65000) spectroscopy. A detailed introduction for these three instruments of the Xinglong 2.16-m reflector can be found in Fan et al. (\cite{{2016PASP..128k5005F}}).  \\

Many telescopes and spectrographs in the world have available the efficiency estimations in their websites. For example, the system efficiency of the High Efficiency and Resolution Multi-Element Spectrograph at the 3.9-m Anglo-Australian Telescope (Wampler et al.~\cite{{1977VA.....21..191W}}), it is from 6\% to 8\% for Galactic Archaeology (V = 14), which includes the efficiency of telescope, fiber system, spectrograph, and detector (Sheinis et al.~\cite{{2015JATIS...1c5002S}}). The system efficiencies of the High Resolution Echlle Spectrograph of the Lijiang 2.4-m telescope at Yunnan observatory are $2\%$ for $1.''2$ fiber diameter and  $3\%$ for $2.''0$ fiber diameter\footnote{http://www.gmg.org.cn/v2/detail/instrument/24}. The ESO Faint Object Spectrograph and Camera at the ESO 3.6-m telescope has different grisms\footnote{http://www.ls.eso.org/sci/facilities/lasilla/instruments/efosc-3p6/docs/Efosc2Grisms.html\#grisms}, most of which have system efficiencies around $30\%$. However, for all the 
telescopes mentioned above, the efficiency for the spectrograph with different slits and different 
grisms have not been measured systemically and detailedly. According to the statistics in the recent 
10 years, the most frequently used instrument (almost up to half of the 2.16-m reflector's observing time) is the BFOSC.  \\

In this work, we investigate the total efficiency of the telescope with BFOSC mounted on, and make some suggestions for the telescope users and technicians. This paper is organized as follows: Sect.~\ref{xinglong} is an introduction of the Xinglong Observatory; in Sect.~\ref{observation} and Sect.~\ref{data}, the observations and data reduction are presented, respectively; in Sect.~\ref{impacts}, we discuss the impacts of slit widths, weather conditions, stellar brightness, and mirror reflectivities on the efficiency estimations; finally a brief summary is given in Sect.~\ref{conc}. 
\section{The Xinglong Observatory of NAOC}

\label{xinglong}
The Xinglong Observatory (XO) of National Astronomical Observatories, Chinese Academy of Sciences 
(NAOC), with the coordinates: \ang{40;23;39} N, \ang{117;34;30} E, was founded in 1968. Until now, it is the largest optical astronomical observatory in the Asian continent. There are nine telescopes, with effective aperture greater than 50-cm located at the XO. The average altitude of the XO is $\sim$ 960 m, and it is located at Xinglong county, Chengde city, Hebei province, which is $\sim$ 120 km northeast of Beijing. The mean and median seeing values of the XO are around $1''.9$ and $1''.7$, respectively, with over one year statistics. For most of the time, the mean and median values of wind speed of the XO are ranging from 1 m s$^{-1}$ to 3.5 m s$^{-1}$, and the sky brightness at the zenith is around 21.1 mag arcsec$^{-2}$ (V-band). About 63\% of the nights per year can be used for spectroscopic observations based on the statistics of observational data in 2007 -- 2014. More detailed introduction of the observing conditions of the XO can be found in Zhang et al. (\cite{{2015PASP..127.1292Z}}).

\section{Observations and Data Reduction}
\label{obs and data}

\subsection{Observations}
\label{observation}

In this work, we chose two flux standard stars with different brightness from the list of the European Southern Observatory (ESO) spectrophotometric standard stars\footnote{http://www.eso.org/sci/observing/tools/standards/spectra.html}. Table \ref{targets} lists the information of the two standard stars, where the HD93521 is $\sim$ 4 
mag brighter than the Feige34 in V-band. Since the efficiency varies much with different slit widths and grisms, we can investigate the variation of efficiencies by observing standard stars with different brightness. The spectral types of the two standard stars are both O-types. Therefore, we can eliminate the effects on the efficiency estimations due to the different spectral types of observing targets. Table \ref{slits} presents the parameters of the slits of the BFOSC from Huang et al. (\cite{{2012..April...1}}). Table \ref{bfosc} presents the parameters of the grisms/prisms/echelle for the BFOSC (see Fan et al.~\cite{2016PASP..128k5005F} for details). The long slits with the length of $9''.4$ in Table \ref{slits} are used for the common grisms (No. 2--10 in Table \ref{bfosc}) spectroscopic observations, and the short slits are used for the echelles (No. 11--13 in Table \ref{bfosc}) spectroscopic observations. The echelle E13 is designed for measuring the velocity field of extended sources with the third order spectrum, and the V-band filter is used to remove the other 
orders of the spectrum (Fan et al. \cite{{2016PASP..128k5005F}}). \\

\begin{table*}
\centering 
\caption{The Information of the Two Observed Standard Stars, including the Coordinates, the V Magnitudes and the Spectral Types.
 \label{targets}}

\begin{tabular}{lccrc}

\hline\hline
Name    &  RA  & Dec & V (mag)  &  Spectral type     \\
\hline                                                                                        
HD$93521$    &   10:48:23.51 & +37:34:12.8 &  7.04 & O       \\
\hline                                                                                        
Feige34    & 10:39:36.71 &+43:06:10.1    & 11.18 &  O        \\
\hline                                                                                        
\end{tabular}
\end{table*}

\begin{table}
\caption{The Parameters of the Slits of the BFOSC. 
 \label{slits}}
\centering

\begin{tabular}{lc|cc}

\hline\hline
 Long Slits  &          &     Short Slits   &       \\                                                                                       
\hline                                                                                        
   Width      &   Length     &   Width      &   Length   \\
\hline                                                                                        
  ($''$)    &  ($'$)    & ($''$)    &  ($'$)  \\
\hline
$0.6$      &     $9.4$      &$0.6$  & $3.5$     \\
\hline
$0.7$      &      $9.4$      &$1.0$  & $4.0$     \\
\hline
$1.1$      &      $9.4$      &$1.6$  & $3.6$       \\
\hline
$1.4$     &      $9.4$      &$2.3$  & $3.7$      \\
\hline         
$1.8$      &      $9.4$      &$3.2$  & $3.7$      \\
\hline
$2.3$      &      $9.4$      &       &    \\
\hline
$3.6$      &      $9.4$      &       &    \\
\hline
$7.0$      &      $9.4$      &       &  \\
\hline                                                                                                      
$14.0$     &     $9.4$      &        &   \\
\hline                                                                                                      

\end{tabular}
\end{table}
\begin{table*}  
\caption{The Parameters of the Grisms/Prisms/Echelle for
the BFOSC. 
 \label{bfosc}}
\centering
\begin{tabular}{lccccc}

\hline\hline
Number &Name     &  Spec.Ord. & Recip. Disp. & Sp. Res. Per Pixel. & Wav. Range   \\                                                                                       
         &       & (m)        & (\AA/mm)      &  (\AA/pix)          & (\AA)    \\
\hline                                                                                        
1 &P1      &            &$573$ -- $2547$ & $8.6$ -- $38.2$    & $4000$ -- $5600$ \\
\hline
2&G3      &     1      &$139$  & $3.12$     & $3300$ -- $6600$ \\
\hline
3&G4      &     1      &$198$  & $4.45$     & $3600$ -- $8700$ \\
\hline
4&G5      &     1      &$199$  & $4.47$     & $5200$ -- $10000$ \\
\hline
5&G6      &     1      &$88$  & $1.98$     & $3300$ -- $5450$ \\
\hline         
6&G7      &     1      &$95$  & $2.13$     & $3780$ -- $6760$ \\
\hline
7&G8      &     1      &$80$  & $1.79$     & $5800$ -- $8280$ \\
\hline
8&G10      &     1      &$392$  & $8.80$     & $3300$ -- $10000$ \\
\hline
9&G11      &     1      &$295$  & $6.63$     & $3600$ -- $9600$ \\
\hline                                                                                                      
10&G12      &     1      &$837$  & $18.8$     & $5200$ -- $10000$ \\
\hline                                                                                                      
11&E9+G10      &  $22$--$10$  &$16.8$ -- $38.4$ & $0.38$ -- $0.86$  & $3300$ -- $10000$ \\
\hline
12&E9+G11      & $18$--$9$ &$21.0$ -- $47.9$ & $0.47$ -- $1.076$    & $3900$ -- $9800$ \\
\hline
13&E9+G12      & $12$--$6$ &$29.0$ -- $73.2$ & $0.65$ -- $1.64$    & $5200$ -- $10000$ \\
\hline
14&E13+V      &     $3$       &$33.1$  & $0.76$    & $4980$ -- $5990$ \\
\hline
\end{tabular}
\end{table*}

For the long slits of the BFOSC, most users choose the slit width of $1''.8$ or $2''.3$, 
depending on the real-time seeing during the observations. While for the short slits of the BFOSC, 
most observers choose slit width of $1''.6$ or $2''.3$. In this work, we choose the slit widths of $1''.8$, $2''.3$, $7''.0$ and $14''.0$ for the long slits; and choose the slit widths of $1''.6$ and $2''.3$ for the short slits. The observation strategy is that we observe the two standard stars with all the grisms/prisms/echelle of the BFOSC with these slit widths. For the E13+V, we observe the two standard stars with slit widths of $1''.8$, $7''.0$, and $14''.0$. However, due to the limited time, for the Feige34, the slit width of $7''.0$ for the E13+V and E9+G12 are not observed this time. The P1 is 
a straight prism, which is seldom used, we do not observe.\\

For all of the observations, the spectral quality is high, with signal-to-noise ratio (SNR) $\geq 100$. 
Table \ref{obs} presents the observation dates. We observed the two ESO standard 
stars on February $19$th, March $8$th, and March $9$th, $2017$, with the long slits and short slits 
of the BFOSC. The seeing was $\sim$ $3''.0$ during the observations. For the grisms G$4$ and G$7$, we also used the 385LP filter for removing the 2nd-order spectrum of wavelength $\geq 385$ nm. 

\begin{table*} 

\caption{The Observation Dates of Grisms.
 \label{obs}}
 \centering 
\begin{tabular}{l l}
\hline\hline
Date    &  Grism \\
\hline                                                                                        
2017-02-19             &  G3, G4, G5, G6, G7, G8       \\
\hline                                                                                        
 2017-03-07            &  G10, G11, G12, E13+V\\
\hline                                                                                        
  2017-03-08           &  E9+G10, E9+G11, E9+G12\\  
\hline
\end{tabular}
\end{table*}

\subsection{Data Reduction}
\label{data}

The raw data was processed with the standard procedure for data reduction, with commands in Image 
Reduction and Analysis Facility (IRAF)\footnote{IRAF is distributed by the National Optical Astronomy Observatory, 
which is operated by the Association of Universities for Research in Astronomy, Inc.(AURA) under cooperative 
agreement with the National Science Foundation} and Interactive Data Language (IDL). For the CCD used by the BFOSC, 
the dark current is $<$ 0.002 e$^-$/pixel/s ($-85 ^\text{o}$C). We did the bias and flat-field corrections for 
object images with the IRAF task $ccdproc$. The IRAF task $dispcor$ was used for wavelength calibration, and 
obtained the total number of ADU (Analog-to-Digital Units) in different wavelength, which is represented as F$_{adu}
$ in Equation \ref{eff}. We downloaded the calibrated spectra of the two standard stars HD$93521$ and Feige$34$ from 
the ESO website. Since the wavelength interval from the  ESO spectra does not match with that of the BFOSC wavelength well, we use the IDL command $interpol$ to interpolate the ESO wavelength to our system. \\

In order to calculate the total 
efficiency of the system, many factors need to be considered, including the atmospheric extinction, the reflectivity of the primary and secondary mirrors, the transmissions of 
grisms and the quantum efficiency of the CCD. The total efficiency of the system can be estimated 
through observing standard stars (see Fan et al.~\cite{2016PASP..128k5005F}) as follows,
\begin{equation}
\label{eff}
 \eta(\lambda)= \frac{F_{adu} \cdot Gain}{F_\lambda \cdot \delta\lambda \cdot S_{tel}}
\end{equation}
where, $F_{adu}$ is the observed number counts of a standard star per second $
(\rm ADU \cdot s^{- 1})$; $Gain$ is the gain of the 
CCD ($\rm e^{- 1} \cdot ADU^{- 1}$); $ F_\lambda$ is the theoretical photon flux of a standard 
star, and we can derive it from its AB mag $(\rm photon \cdot s^{- 1} \cdot cm ^{- 2} \cdot \AA^{- 1})$; $\delta\lambda$ is the dispersion of the grating for spectroscopic observations (\AA); $S_{tel}$ is the effective area of the primary mirror of the telescope
 $(\rm cm^2)$; and $\lambda$ is the the wavelength (\AA).\\ 
\section{The Impacts on the Efficiency Estimations}
\label{impacts}

\subsection{The Impact of Slit Widths on the Efficiency Estimations}
\label{impact of slit width}
We used the Equation \ref{eff} to calculate the total efficiency of the telescope with BFOSC, with the different slit widths of different grisms/prisms/echelle. In fact, there are some absorption lines in the efficiency curves, including the gaseous atmosphere absorption lines (e.g. O$_2$, H$_2$O), the stellar atmosphere absorption lines (e.g. H$\alpha$, H$\beta$) and the noise. In order to show the efficiency curves more clearly, we smooth all these atmosphere absorption lines.                        \\

\begin{figure}

  \begin{minipage}[t]{0.495\linewidth}
  \centering
   \includegraphics[width=79mm]{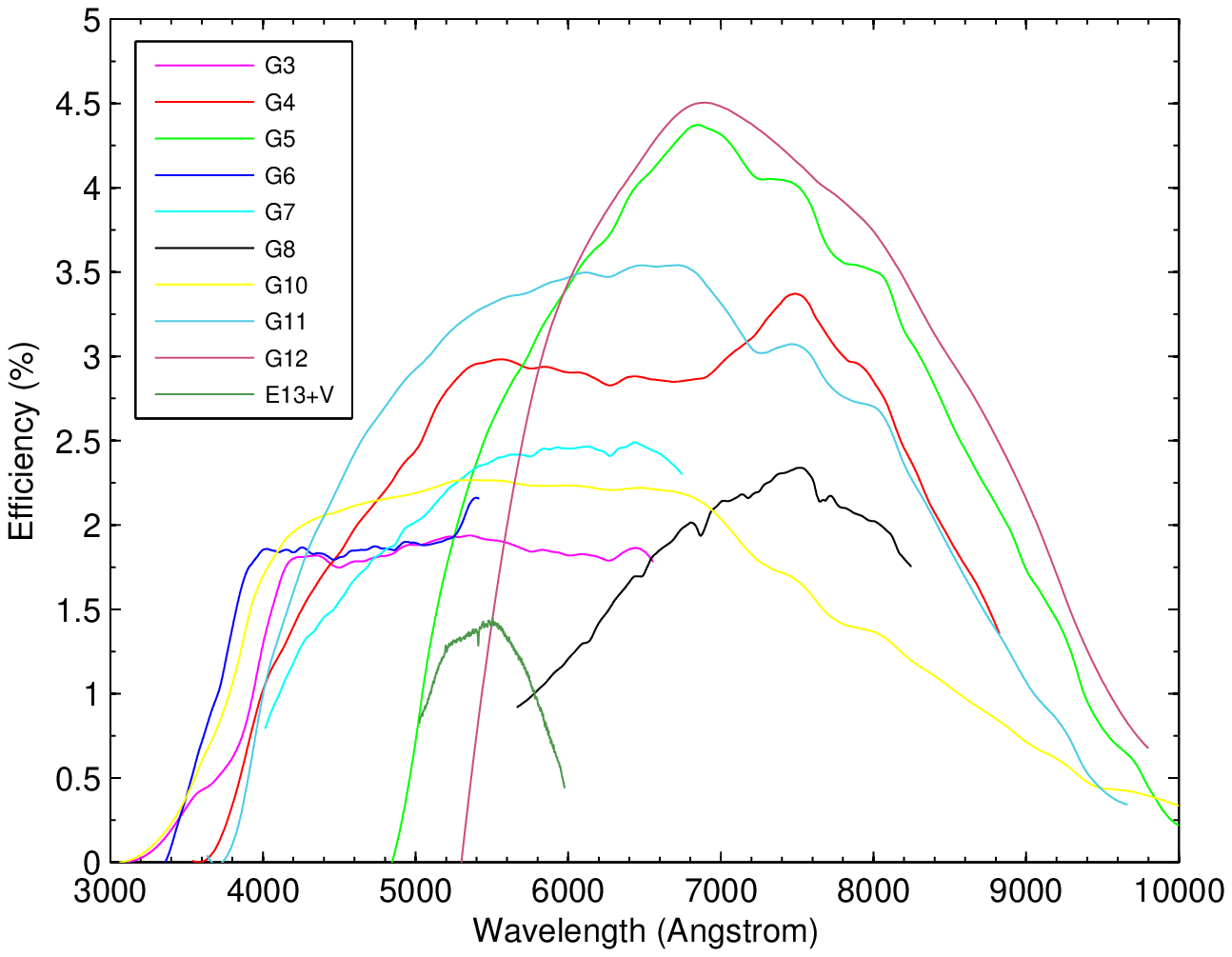}
   
  \end{minipage}%
  \begin{minipage}[t]{0.495\textwidth}
  \centering
   \includegraphics[width=79mm]{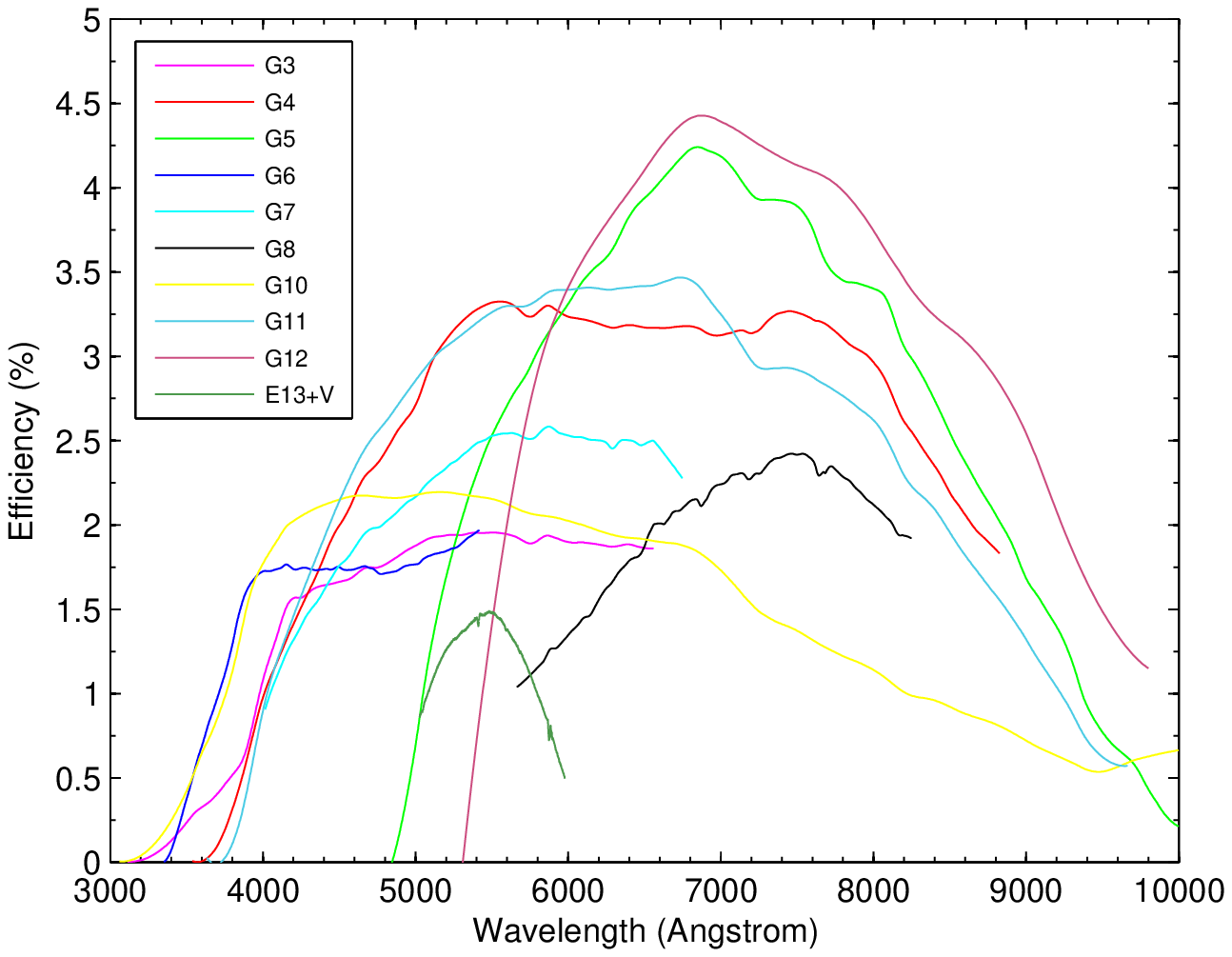}
 
  \end{minipage}%
 
  \caption{{The total efficiencies of the different grisms of the BFOSC, estimated from the observations of the Feige34 (left panel) and HD93521 (right panel) with a slit width of $1''.8$. The different colored lines represent the different grisms.} }
   \label{fig18}
\end{figure}

\begin{figure}

  \begin{minipage}[t]{0.495\linewidth}
  \centering
   \includegraphics[width=79mm]{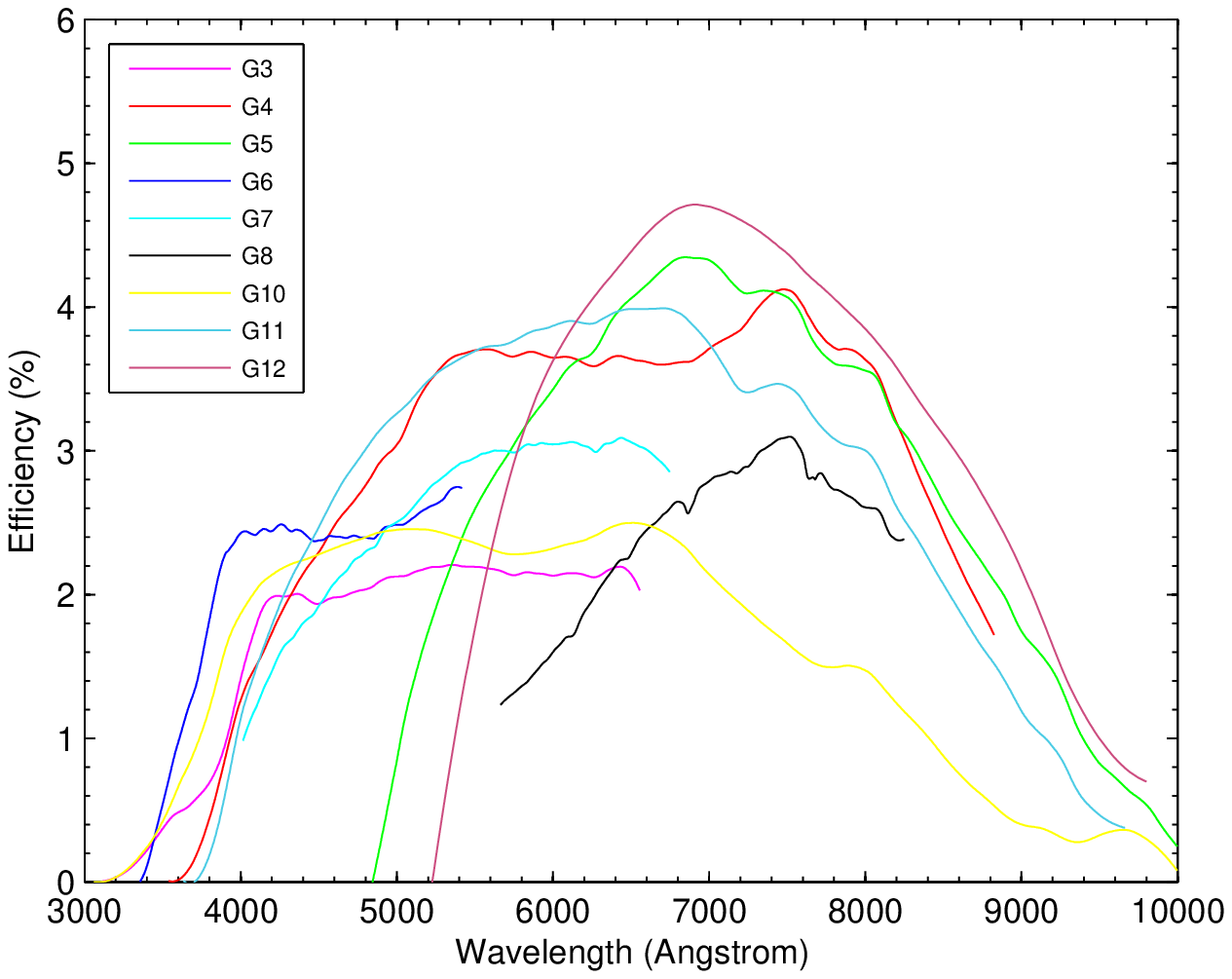}
   
  \end{minipage}%
  \begin{minipage}[t]{0.495\textwidth}
  \centering
   \includegraphics[width=79mm]{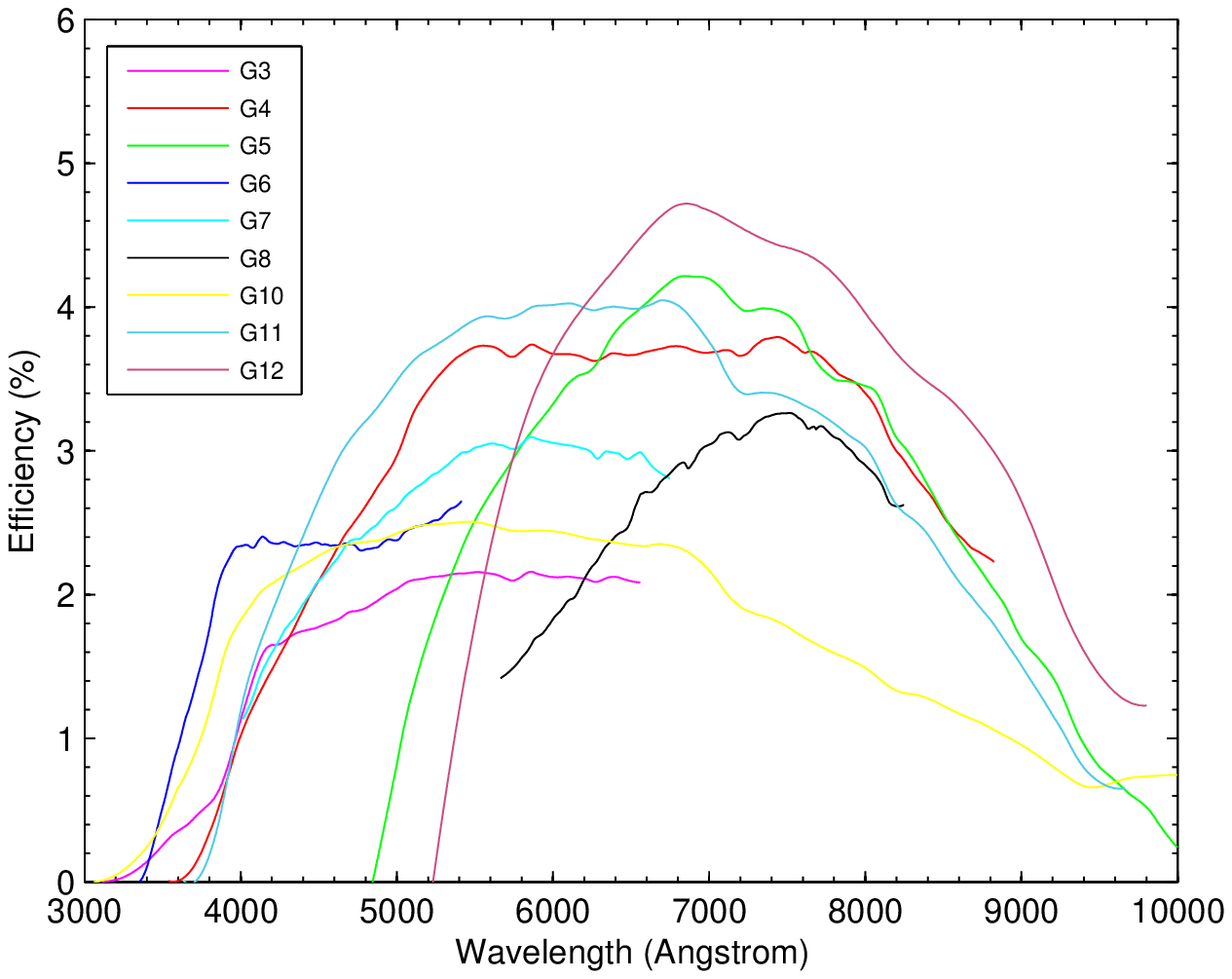}
  \end{minipage}%
  \caption{{Same as Fig. \ref{fig18}, but with a slit width of $2''.3$.} }
  \label{fig23}
\end{figure}

\begin{figure}

  \begin{minipage}[t]{0.495\linewidth}
  \centering
   \includegraphics[width=79mm]{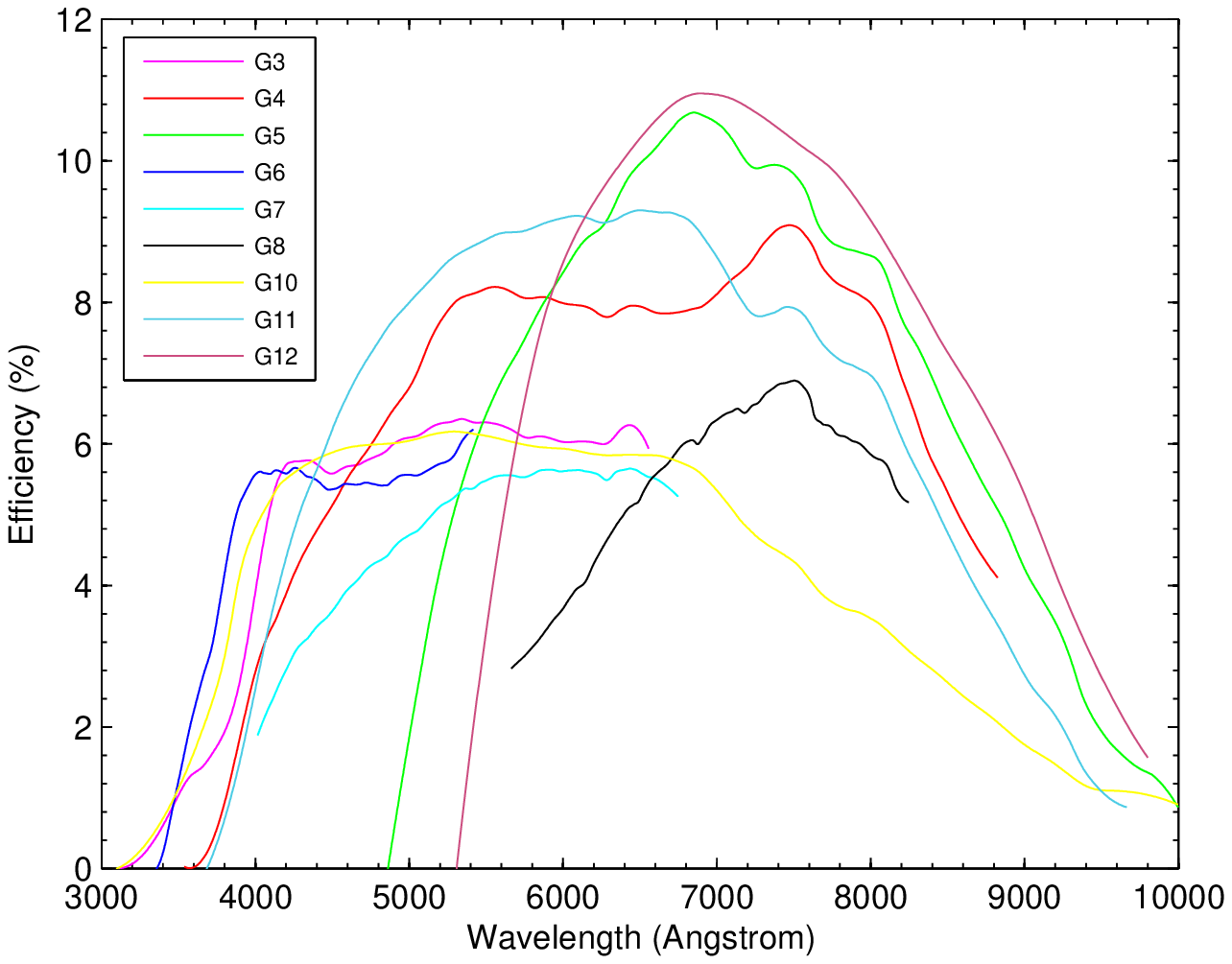}
   
  \end{minipage}%
  \begin{minipage}[t]{0.495\textwidth}
  \centering
   \includegraphics[width=79mm]{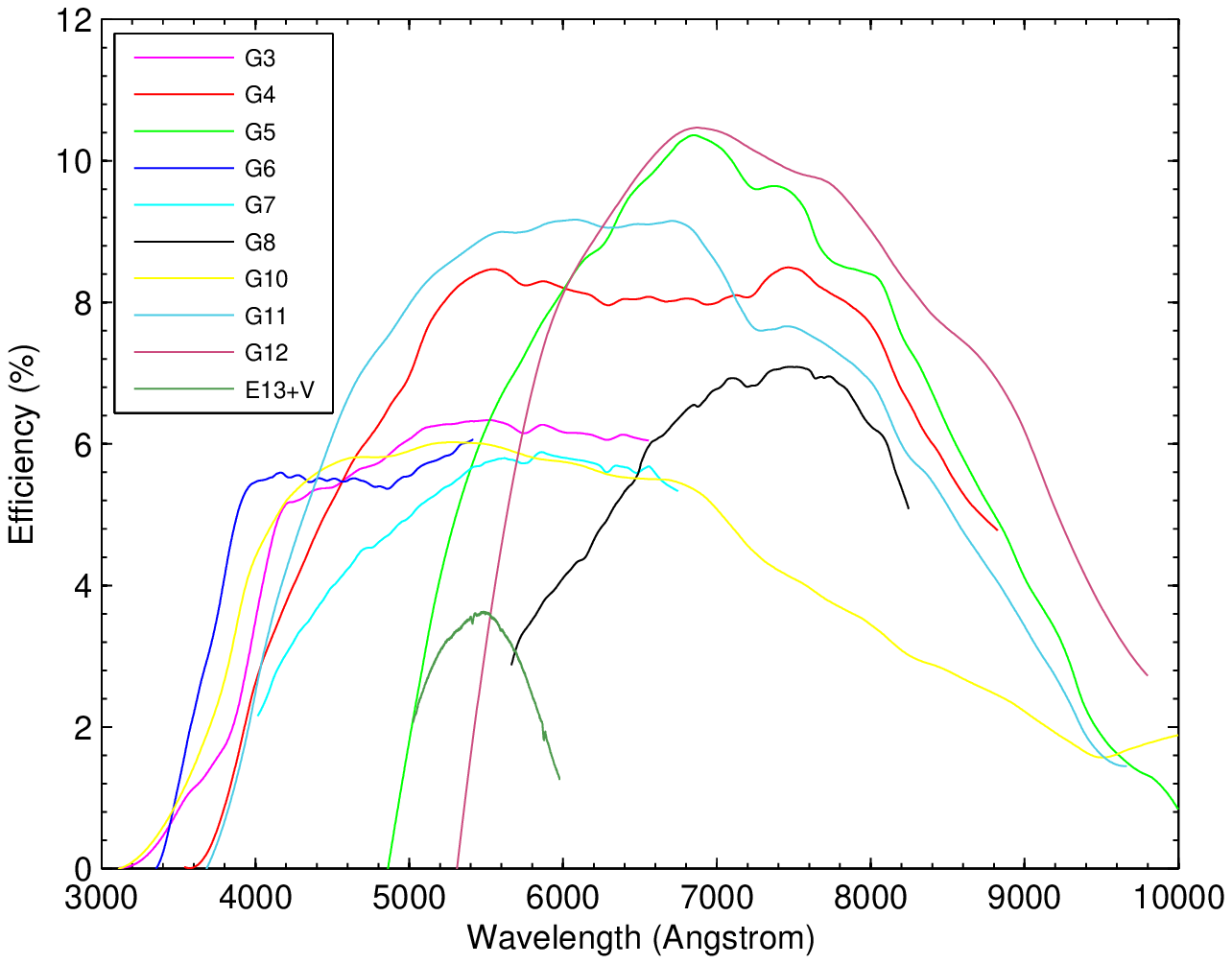}
 
  \end{minipage}%
  \caption{{Same as Fig. \ref{fig18}, but with a slit width of $7''.0$.} }
   \label{fig7}
\end{figure}

\begin{figure}

  \begin{minipage}[t]{0.495\linewidth}
  \centering
   \includegraphics[width=79mm]{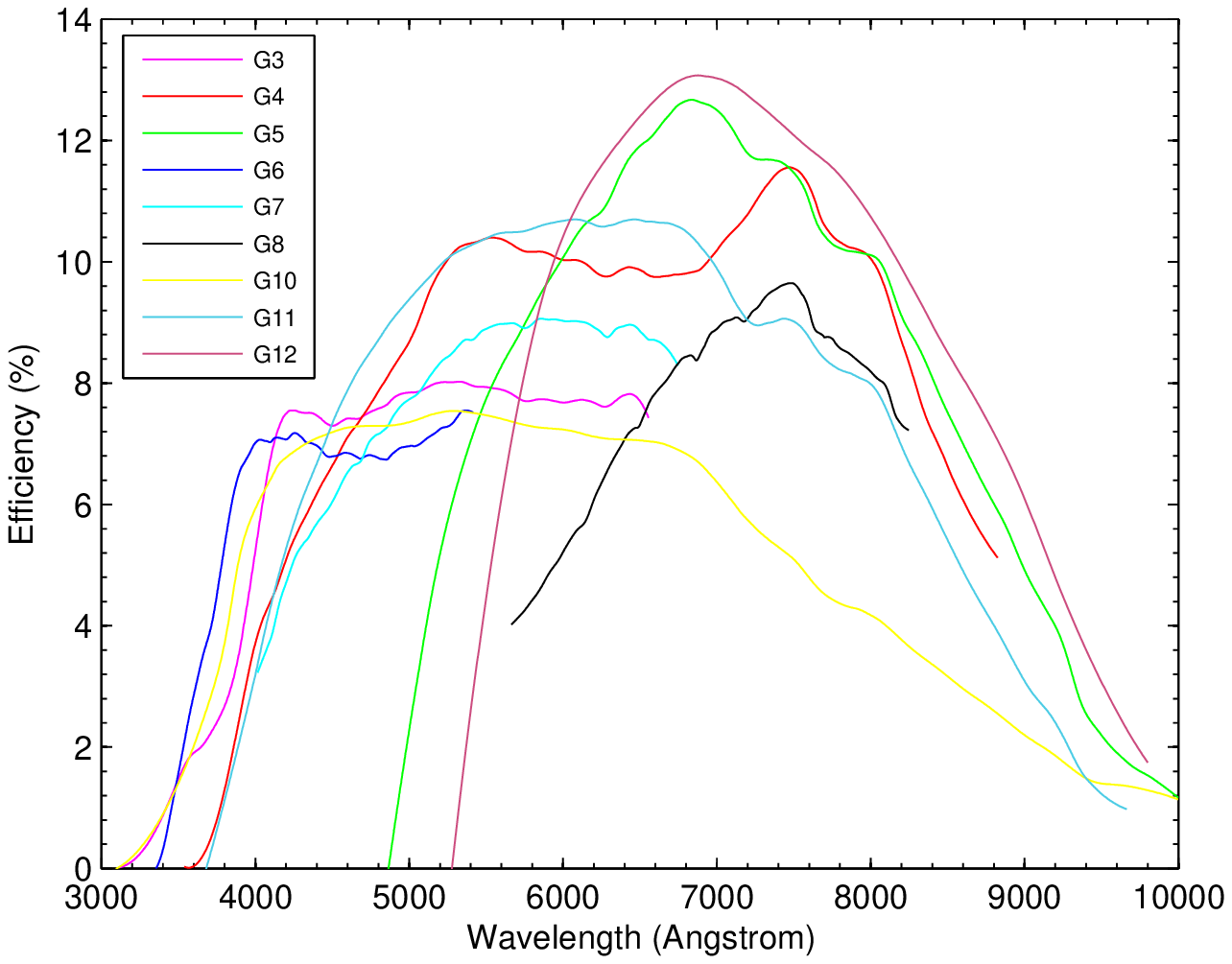}
   
  \end{minipage}%
  \begin{minipage}[t]{0.495\textwidth}
  \centering
   \includegraphics[width=79mm]{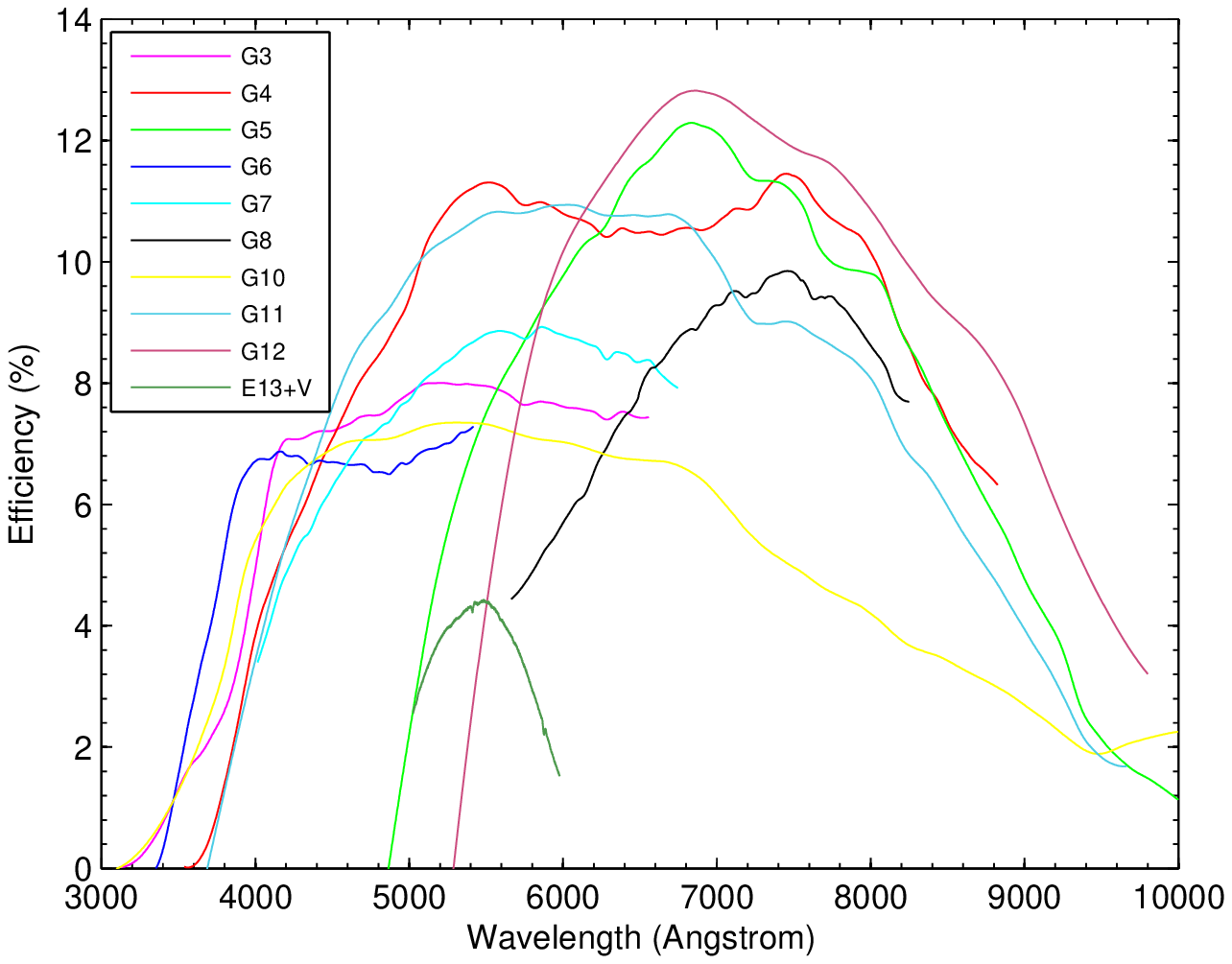}
  
  \end{minipage}
  \caption{{Same as Fig. \ref{fig18}, but with a slit width of $14''.0$.} }
   \label{fig14}
\end{figure}

\begin{figure}

  \begin{minipage}[t]{0.495\linewidth}
  \centering
   \includegraphics[width=79mm]{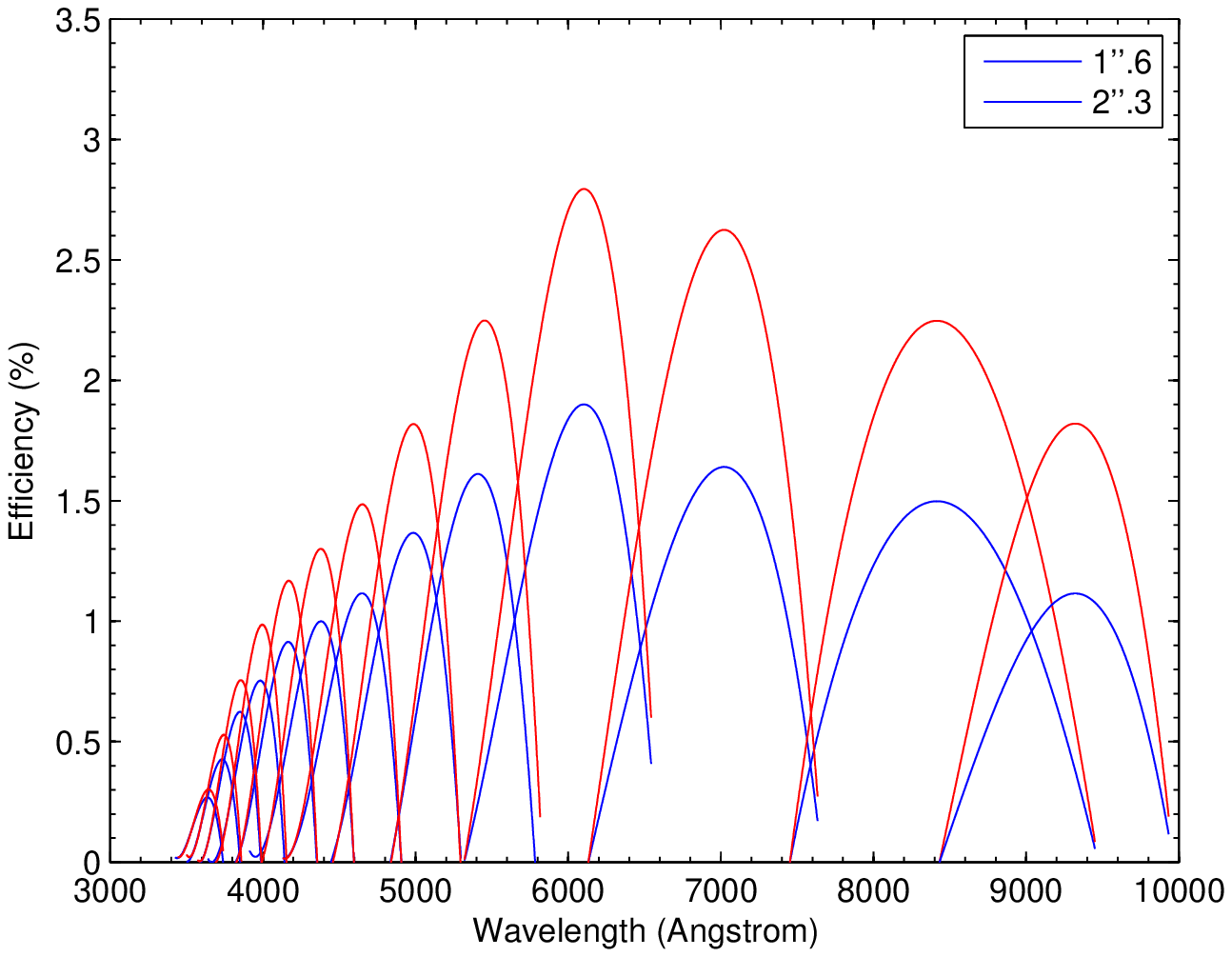}
   
  \end{minipage}%
  \begin{minipage}[t]{0.495\textwidth}
  \centering
   \includegraphics[width=79mm]{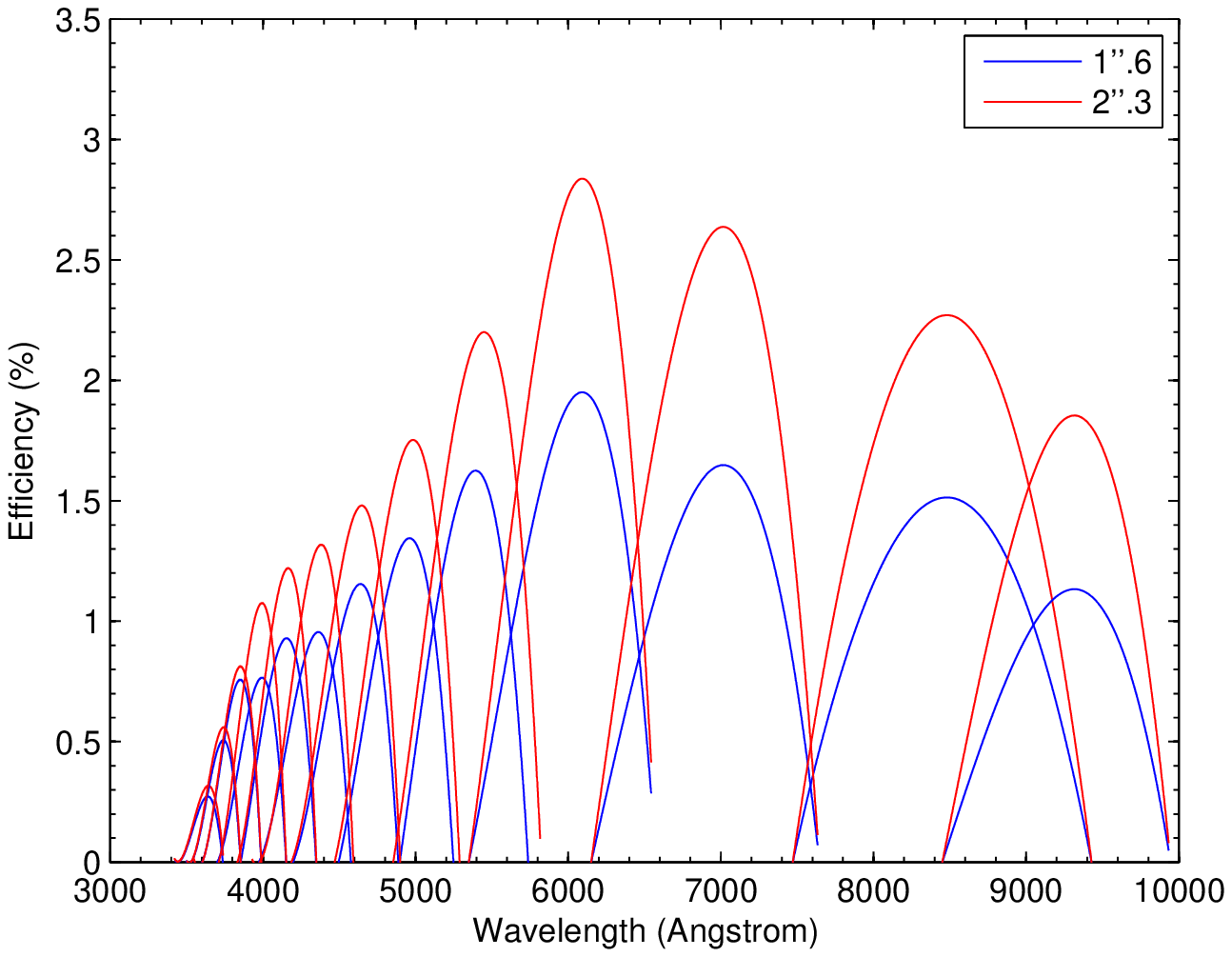}
  \end{minipage}%
  \caption{{The total efficiencies of the E9+G10, estimated from the observations of the Feige34 (left panel) and HD93521 (right panel), with the slit widths of $1''.6$ (blue lines) and $2''.3$ (red lines) , respectively.} }
  \label{fige9g10}
\end{figure}

\begin{figure}

  \begin{minipage}[t]{0.495\linewidth}
  \centering
   \includegraphics[width=79mm]{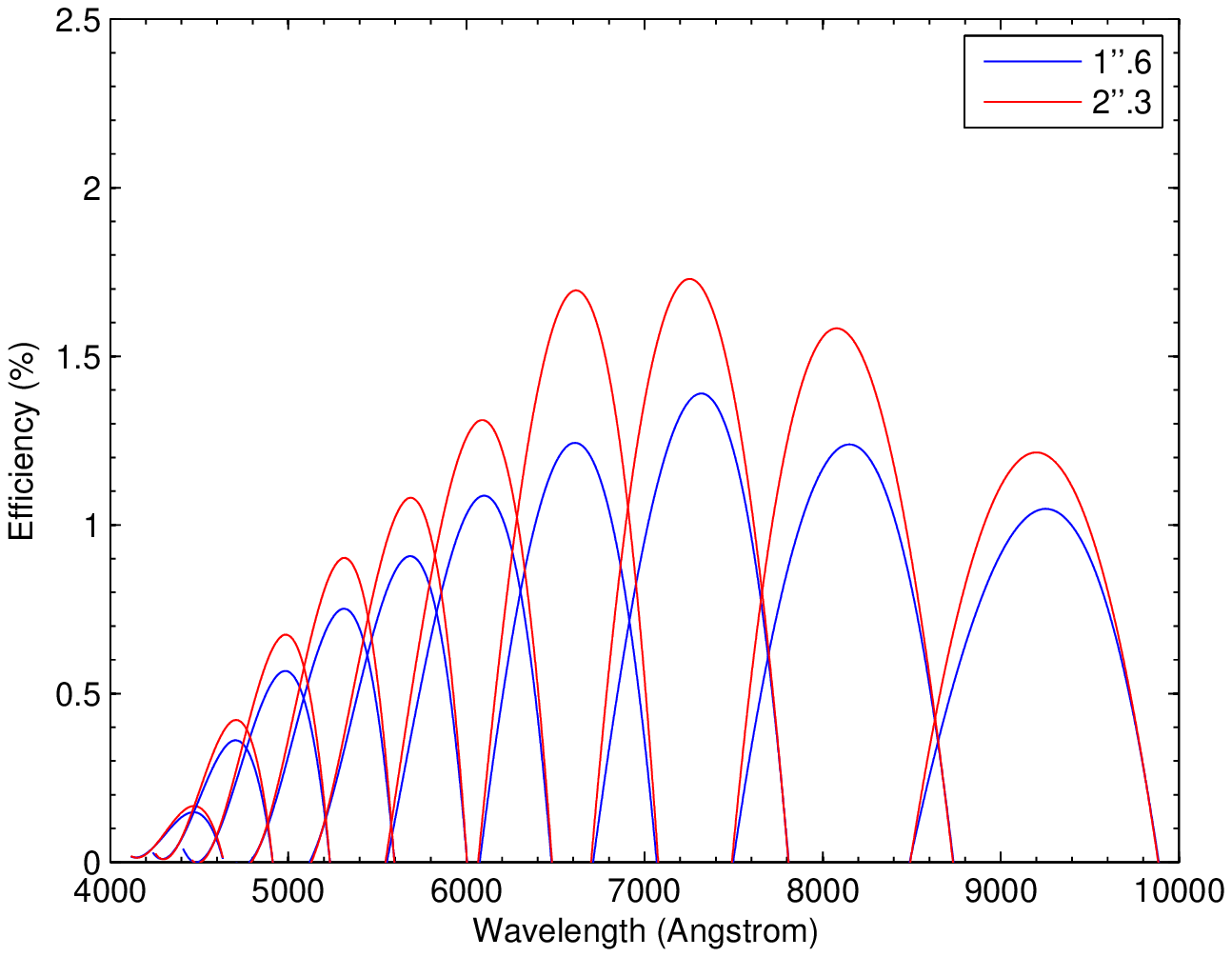}
   
  \end{minipage}%
  \begin{minipage}[t]{0.495\textwidth}
  \centering
   \includegraphics[width=79mm]{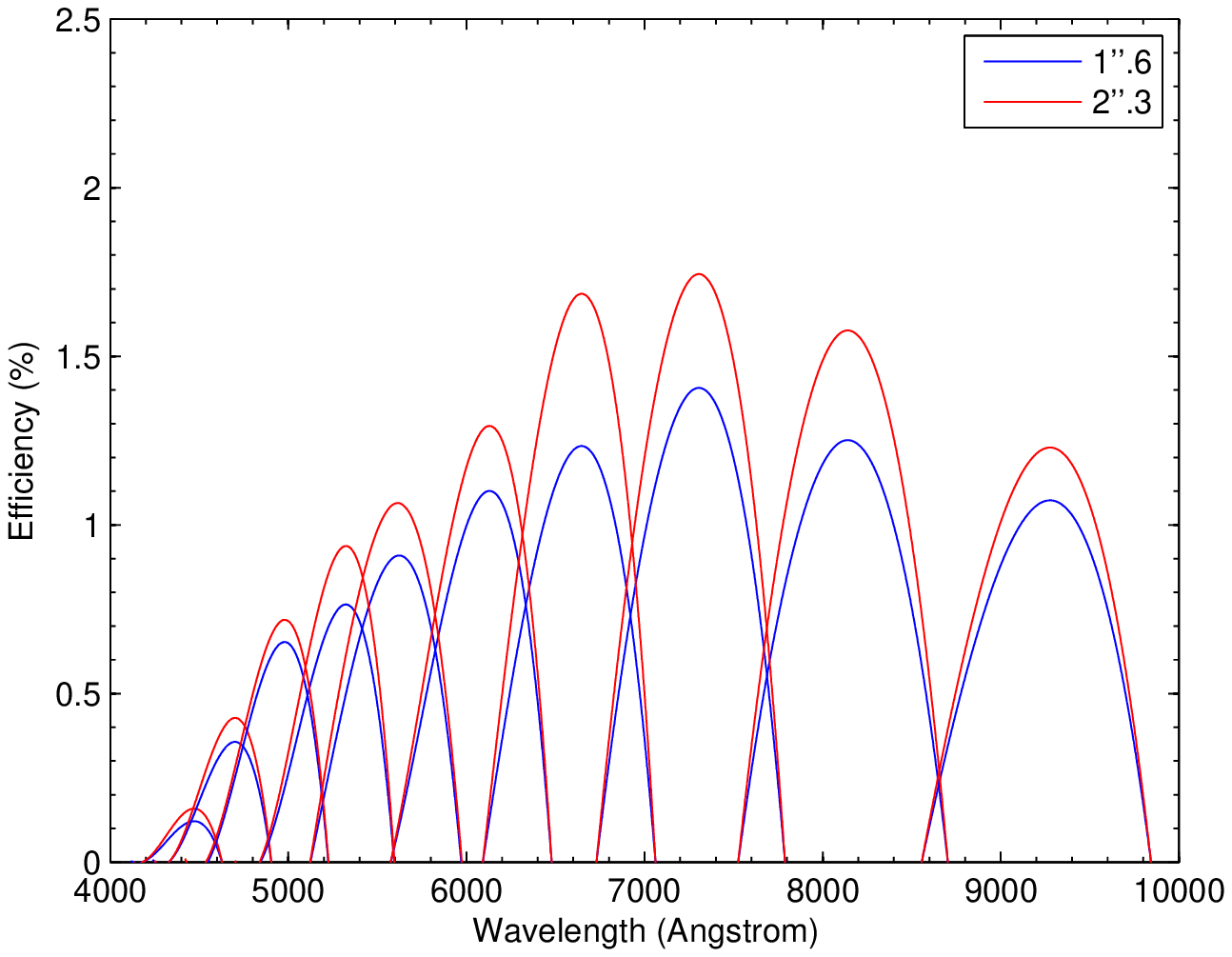}

  \end{minipage}
  \caption{{Same as Fig. \ref{fige9g10}, but with the E9+G11.} }
  \label{fige9g11}
\end{figure}

\begin{figure}
\centering
\includegraphics[width=10cm]{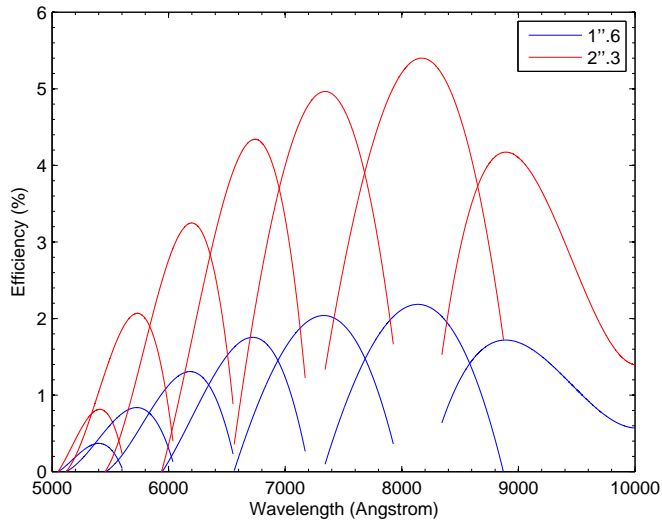}
\caption{The total efficiencies of the E9+G12, estimated from the observations of the HD93521, with the slit widths of $1''.6$ (blue lines) and $2''.3$ (red lines) respectively.}
\label{fige9g12}
\end{figure}

Figures \ref{fig18} -- \ref{fig14} show the total efficiencies of the Feige$34$ 
and HD$93521$ in the wavelength from $\sim$ 3000 \AA \ to $\sim$ 10000 \AA \ for the grisms G3/G4/G5/G6/G7/G8/G10/G11/G12/E13+V, with the slit widths of $1''.8$, $2''.3$, $7''.0$, and $14''.0$, respectively. For the slit width of $1''.8$, the peak of total efficiencies are $\sim$ 1.7\% -- 4.5\%. For the slit width of 
$2''.3$, the peak of the total efficiencies are $\sim$ 2\% -- 5\%. For the slit width of $7''.0$, the peak of the total efficiencies are $\sim$ 4.8\% -- 11\%. For the slit width of $14''.0$, the peak of the total efficiencies are $\sim$ 6.6\% -- 13\%. Figures \ref{fige9g10} and \ref{fige9g11} are present the results of the Feige$34$ and HD$93521$ at different wavelength from $\sim$ 3000 \AA \ to $\sim$ 10000 \AA \ with the slit widths of $1''.6$ and 
$2''.3$, for the E9+G10, E9+G11, respectively. Figure \ref{fige9g12} is the results of the HD$93521$ at different wavelength from $\sim$ 5200 \AA \ to $\sim$ 10000 \AA \ with the slit widths of $1''.6$ and $2''.3$, for 
the E9+G12. 
For the slit width of $1''.6$, in Figure \ref{fige9g10}, the 
peak of the total efficiencies are $\sim$ 0.5\% -- 2.0\%, for the wavelength $\ge 4000$ \AA; In Figure \ref{fige9g11}, 
the peak of the total efficiencies are $\sim$ 0.5\% -- 1.4\%, for the wavelength $\ge 5000$ \AA; In Figure 
\ref{fige9g12}, the peak of the total efficiencies are $\sim$ 0.4\% -- 2.0\%, for the wavelength $\ge 5400$ \AA. For 
the slit width of $2''.3$, in Figure \ref{fige9g10}, the peak of the total efficiencies are $\sim$ 1.0\% -- 3.8\%, for 
the wavelength $\ge 4000$ \AA; In Figure \ref{fige9g11}, the peak of the total efficiencies are $\sim$ 0.6\% -- 1.7\%, for 
the wavelength $\ge 5000$ \AA; In Figure \ref{fige9g12}, the peak of the total efficiencies are $\sim$ 0.8\% -- 5.4\%, for the wavelength $\ge 5400$ \AA.\\

\subsection{The Impact of Stellar Brightness on the Efficiency Estimations}
\label{impact of stellar}

\begin{figure}
\centering
\includegraphics[height=18cm]{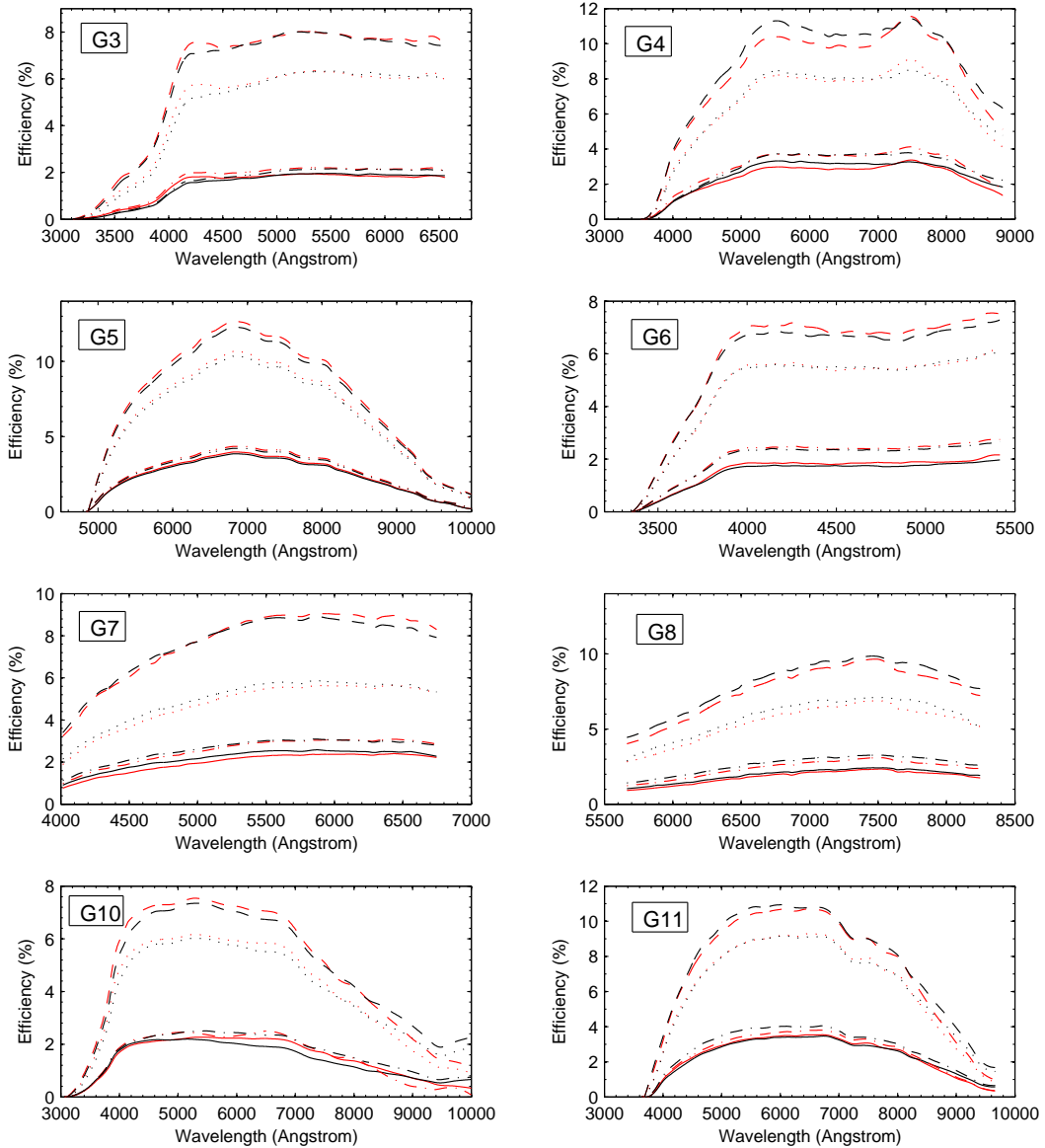}

\caption{The results of the grisms G$3$/G$4$/G$5$/G$6$/G$7$/G$8$/G$10$/G$11$ with the Feige34 (red color) and HD93521 (black color), with the slit widths of $1''.8$ (solid), $2''.3$ (dotted-dashed), $7''.0$ (dotted) and $14''.0$ (dashed), respectively.}

\label{comparetotal}

\end{figure}

Since the total efficiency estimations may depend on the brightness of the observational stars, we also compare the total efficiencies obtained from the observations of the Feige$34$ (the faint one) and HD$93521$ (the bright one). Figure \ref{comparetotal} compare the results of the grisms G$3$/G$4$/G$5$/G$6$/G$7$/G$8$/G$10$/G$11$ with the Feige34 and HD93521, with the slit widths of $1''.8$, $2''.3$, $7''.0$, and $14''.0$, respectively. 
As shown in Figure \ref{comparetotal}, we can clearly find the difference of total efficiencies when choosing different slit widths. For the same grism, the total efficiencies rise with increasing slit widths. However, we do not find a clear relationship between different brightness of stars and efficiency estimations in this work. \\

\subsection{The Impact of Weather Conditions on the Efficiency Estimations}
\label{impact of seeing}

\begin{figure}
\centering
\includegraphics[width=10cm]{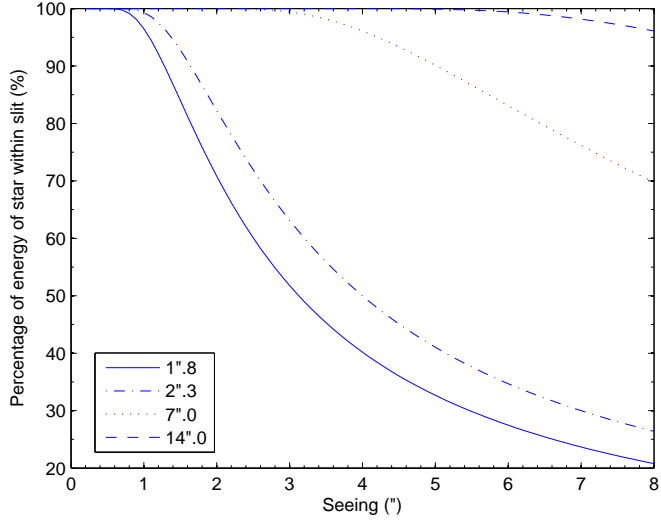}
\caption{The relationship between the percentage of energy of star within the slit and the different seeings. The solid line, dotted-dashed line, dotted line and dashed line are represent the slit widths of $1''.8$ , $2''.3$, $7''.0$ and $14''.0$, respectively.} 
\label{seeing}
\end{figure}

We investigate the weather conditions (seeing) influence on the efficiency estimations of the telescope with BFOSC. 
In this work, we use the full width at half maximum (FWHM) of the star imaging profile, obtained by the Gaussian fitting, to estimate the influence of seeing. Figure \ref{seeing} shows the relationship between the percentage of energy of star within the slit and the different seeings. At Xinglong Observatory, the mean seeing value is $1''.9$ (Zhang et al.~\cite{{2015PASP..127.1292Z}}), therefore, when do the simulation, the seeing we adopt is $2''.0$. We numerically simulate the Gaussian FWHM with slit widths  of $1''.8$ and $2''.3$, and obtain the energy within the slit widths: 73.70\% for 
$1''.8$, and 82.64\% for $2''.3$. With the same method, for the seeing of $3''.0$, we obtain the energy within the slit widths: 54.43\% for $1''.8$, and 63.35\% for $2''.3$. As shown in Figure \ref{seeing}, we can clearly find the seeing significantly affects the percentage of energy of star within the slit, especially for the slit widths of $1''.8$ and $2''.3$. Thus, the weather conditions also influence the efficiency of the telescope and spectrograph, depending on the slit width chosen during the observations.\\

\subsection{The Impact of Mirror Reflectivities on the Efficiency Estimations}
\label{impact of mirror}

\begin{figure}
\centering
\includegraphics[width=10cm]{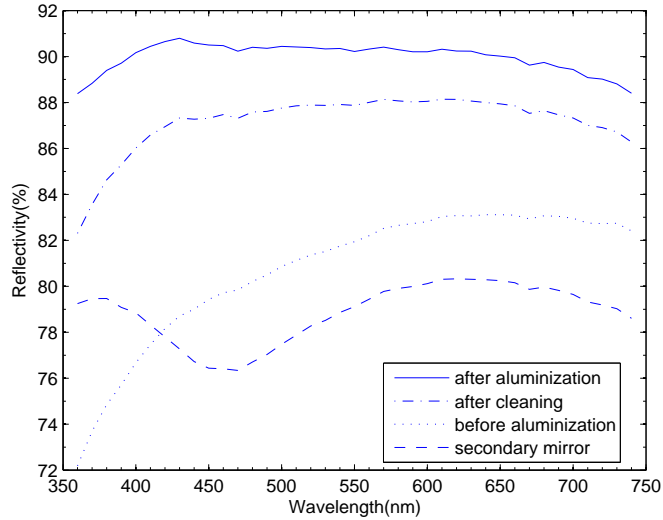}
\caption{The reflectivity curves of the primary and secondary mirrors of the 2.16-m reflector. The solid line, dotted-dashed line, and dotted line are the reflectivity curves obtained after aluminization, after cleaning (before aluminizztion) and before aluminization of the 2.16-m primary mirror, respectively. The dashed line is the reflectivity curve of the secondary mirror.} 
\label{reflectivity}
\end{figure}

It is easy to understand that the reflectivities of the primary and secondary mirrors play an important role in the total efficiency of the telescope with BFOSC. The primary mirror of the 2.16-m reflector is usually aluminized in August or September every year. In 2017, the aluminization time of the primary mirror are during the period of September 4th -- 11th. We use the spectrophotometer CM-2600d\footnote{https://sensing.konicaminolta.us/products/cm-2600d-spectrophotometer/} to measure the reflectivities of the primary and secondary mirrors, which are shown in Figure \ref{reflectivity}. The solid line, dotted-dashed line, and dotted line are represent the reflectivity curves after aluminization, after cleaning (before aluminization) and before aluminization of the 2.16-m primary mirror, respectively. The dashed line is the reflectivity curve of the secondary mirror. As shown in Figure \ref{reflectivity}, we can see that the reflectivity of the primary mirror decreases from $\sim$ 5.99\% (in $\sim$ 7400\AA) to $\sim$ 16.00\% (in $\sim$ 3600\AA) in one year. However, after the cleaning of the primary mirror, the reflectivity usually is improved in the whole wavelength coverage. Since the secondary mirror have been used for ten years, we plan to aluminize it in next work.   \\

\begin{figure}
\centering
\includegraphics[width=10cm]{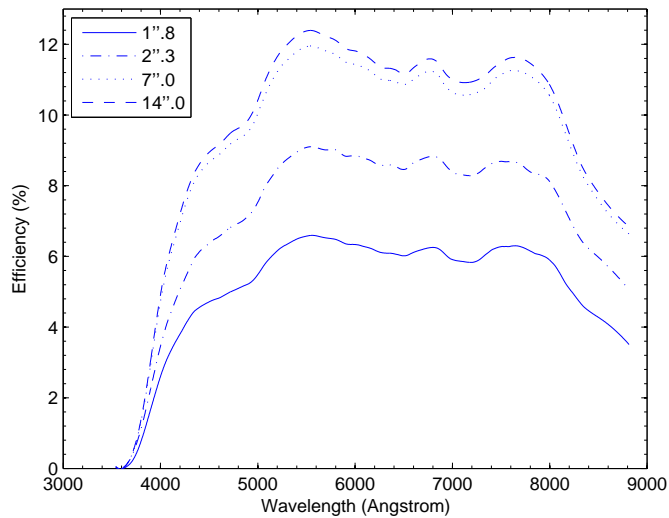}
\caption{The total efficiencies of the grism G4 with different slit widths, after the aluminization of the primary mirror of the 2.16-m reflector. The linetype of solid, dotted-dashed, dotted, and dashed are represent the slit widths of $1''.8$, $2''.3$, $7''.0$ and $14''.0$, respectively.} 
\label{eff_afterdomo}
\end{figure}

 After aluminization of the primary mirror, we measured the total efficiency of the telescope with BFOSC following the method mentioned above. Since the two standard stars HD93521 and Feige34 are not observable at this time, we choose another standard star HR8634 (V = 3.40 mag, spectral type B8V ) from the ESO website. The HR8634 was observed with the grism G4 with slit widths of $1''.8$, $2''.3$, $7''.0$ and $14''.0$ on September 16th, 2017, which are shown in Figure \ref{eff_afterdomo}, in the solid, dotted-dashed, dotted, and dashed lines. The seeing was $\sim$ $2''.0$ during the observations. Compared with Figure \ref{comparetotal} (the grism G4), we can see the peak of total efficiencies have been improved $\sim$ 3.1\%, $\sim$ 5.0\%, $\sim$ 2.9\% and $\sim$ 0.8\% with the slit widths of $1''.8$, $2''.3$, $7''.0$ and $14''.0$, respectively. It is also noted that in the blue-band ($\sim$ 5549 \AA), the total efficiencies have been improved $\sim$ 3.6\%, $\sim$ 5.4\%, $\sim$ 3.7\%, and $\sim$ 2.0\% with the slit widths of $1''.8$, $2''.3$, $7''.0$, and $14''.0$, respectively. While in the red-band ($\sim$ 7656 \AA), the improvement are $\sim$ 2.9\%, $\sim$ 4.9\%, $\sim$ 2.0\%, and $\sim$ 0.5\% with the slit widths of $1''.8$, $2''.3$, $7''.0$, and $14''.0$, respectively. It shows that the aluminization are more important for the improvement of total efficiency in the blue-band ($\sim$ 5549 \AA) than that in the red-band ($\sim$ 7656 \AA). As the seeing was $\sim$ $2''.0$ during the observations of the HR8634, better than that during the observations of the HD93521 and Feige34 ($\sim$ $3''.0$), combined with Figure \ref{seeing}, it shows that the seeing significantly affects the total efficiency of the telescope with BFOSC, especially for the slit widths of $1''.8$ and $2''.3$.


\section{Conclusions}
\label{conc}

We have systematically investigated the total efficiency (including the atmospheric extinction, the reflectivity of the primary and secondary mirrors, the transmissions of 
grisms and the quantum efficiency of the CCD) of different grisms of the BFOSC with different slit widths, and analyzed the factors that possibly impact the efficiency of the telescope and spectrograph, which are 
important for the observations and the operations of the telescope. 
The grism G$12$ has the highest peak of the total efficiency ($\sim$ 13.0\%) of all the grisms of the BFOSC at the wavelength of $\sim$ 6876 \AA. For echelles, the E9+G12 has the highest peak of the total efficiency ($\sim$ 5.4\% at the wavelength of $\sim$ 8160\AA). For wavelength range of 3300 -- 7000 \AA, the total efficiency of the grism G$7$ is the highest ($\sim$ 8.9\% at the wavelength of $\sim$ 5861 \AA); for the wavelength range of 5200 -- 10000 \AA, the total efficiency of the grism G$12$ is the highest ($\sim$ 13.0\% at the wavelength of $\sim$ 6876 \AA); for wavelength range of 3600 -- 8700 \AA, the total efficiency of the grism G$4$ is the highest ($\sim$ 11.5 at the wavelength of $\sim$ 5371 \AA). As the median and mean seeing values of Xinglong Observatory are around $1''.7$ and $1''.9$, respectively (Zhang et al.~\cite{{2015PASP..127.1292Z}}), we recommend the observers to use the slit width of $1''.8$ or $2''.3$. \\

After investigating the reflectivities of the primary and secondary mirrors of the 2.16-m reflector, we suggest that the mirrors should be cleaned and aluminized annually, especially after the pollen season and dust storm in spring, at least once yearly. Comparing the total efficiencies obtained before and after aluminization of the the primary mirror of the 2.16-m reflector, we can see in the blue-band ($\sim$ 5549 \AA), the total efficiencies have been improved $\sim$ 3.6\%, $\sim$ 5.4\%, $\sim$ 3.7\%, and $\sim$ 2.0\% with the slit widths of $1''.8$, $2''.3$, $7''.0$, and $14''.0$, respectively. While in the red-band ($\sim$ 7656 \AA), the improvement are $\sim$ 2.9\%, $\sim$ 4.9\%, $\sim$ 2.0\%, and $\sim$ 0.5\% with the slit widths of $1''.8$, $2''.3$, $7''.0$, and $14''.0$, respectively. It shows that the aluminization is more important for the improvement of total efficiency in the blue-band than that in the red-band. The seeing also significantly affects the total efficiency of the telescope with BFOSC, especially for the slit widths of $1''.8$ and $2''.3$.

From the total efficiencies of the 2.16-m telescope and BFOSC, we can see that the peak of the total efficiencies (including the atmospheric extinction, the reflectivity of the primary and secondary mirrors, the transmissions of the grisms and the quantum efficiency of the CCD) of the 2.16-m reflector and BFOSC are around 6.6\% -- 13.0\%. We will carry out more detailed investigations and improve the efficiency of the telescope and instrument in future work.

\begin{acknowledgements}
 This work is supported by the Open Project Program of the Key Laboratory of Optical Astronomy, National Astronomical Observatories, Chinese Academy of Sciences, and the National Natural Science Foundation of China (NFSC) through grants 11503045, 11373003, National Program on Key Research and Development Project Grant No. 2016YFA0400804, National Key Basic Research Program of China (973 Program) No. 2015CB857002. We acknowledge the support of the staff of the Xinglong 2.16-m telescope, and thank Dr. Meng Zhai and night assistants of the 2.16-m telescope for their kind support with obtaining data. 
\end{acknowledgements}

\label{lastpage}


\end{document}